\documentclass[12pt,a4,fleqn,onecolumn,oneside]{article} %bylo 12pt
\usepackage{amsmath,amsfonts,graphicx,amsbsy}
\usepackage[amssymb]{SIunits}
\usepackage{lineno}
\newcommand{\UPF}{\,\mathcal{O}_{PF}}
\newcommand{\U}{\,\mathcal{O}}
\newcommand{\Uprime}{\,\mathcal{O^{\prime}}}

\begin{document}
\title{ \bf Absolute clock synchronisation and special relativity paradoxes}
\author{\bf Jacek Ciborowski$^1$, Marta W{\l}odarczyk$^2$}
\date{\today}
\maketitle
%\tableofcontents

\vspace{1cm}

%\begin{center}

$^1$ University of Warsaw, Faculty of Physics, Ho\.{z}a 69, 00-681 Warszawa, Poland,
\centerline{cib$@$fuw.edu.pl}

$^2$ University of {\L}\'od\'z, Faculty of Physics and  Applied Informatics, Pomorska 149/153,
\centerline{90-236 {\L}\'od\'z, Poland, m.wlodarczyk@merlin.phys.uni.lodz.pl}

%\end{center}

\vspace{2.0cm}

\noindent
\small{{\bf Abstract:} Solving special relativity paradoxes requires rigorous analysis of event timing, due to relative simultaneity  in consequence of the Lorentz transformation. Since clock synchronisation is a convention in special theory of relativity, instead of the Einstein's procedure one may choose  such that offers absolute simultaneity.
We  present in short the corresponding formalism in one  spatial dimension. We show that paradoxes do not arise
with this choice of synchronisation and  descriptions of these issues are  exceptionally simple and consistent for both observers involved.
 }

\vspace{2cm}
\section{Introduction}\label{intro}
The coordinate time is defined by the procedure of clock synchronisation in a given reference frame~\cite{cite:Reichenbach-1},\cite{cite:grunbaum},\cite{cite:winnie1},\cite{cite:winnie2}.
In the special theory of relativity (STR)~\cite{cite:Einstein-1905}, the Einstein's clock synchronisation procedure  reflects  the fundamental assumption of constancy and isotropy of the velocity of light, $c$,  in any inertial reference frame.   Correctness of this  assumption cannot be verified experimentally since it is not possible to determine a one-way (i.e. open path) velocity of light, other than $c$,  using light signals: the velocity $c$ is  the average velocity of light over a closed path and thus can be determined   using  a single clock without any conventions.

The Einstein's clock synchronisation convention leads to the Lorentz transformation and, in consequence, to the  result that simultaneity is not absolute, i.e. events simultaneous in one reference frame are not simultaneous in another reference frame  in relative motion w.r.t. to the former. After postulating STR, numerous problems have been formulated that possessed apparently  paradoxical character,  including the ladder--barn  paradox, the twin paradox as well as the much later proposed Bell's spaceship paradox. These paradoxes  arise because of relative   simultaneity in STR --  their explanations  require  precise  analyses  regarding timing of events  and  are  non-intuitive.

Since the procedure of synchronising  distant clocks   is a convention,  one has freedom to adopt synchronisation  procedures different from that of Einstein's, provided the velocity of light over closed paths equals $c$. Obviously,  transformation equations would differ  in such cases  from those of the standard  Lorentz transformation but qualitative and quantitative results must remain unaltered. If so,
could one define  a  suitable   procedure  that would  provide  a   more simple and  more intuitive explanation of  STR paradoxes?   Since   the notion of simultaneity  is  the underlying message in this context, it is  natural  to consider   a synchronisation procedure   that offers  absolute simultaneity.
It has been shown by Rembieli\'{n}ski~\cite{cite:Kuba-2},\cite{cite:Kuba-1} that indeed, by  appropriate redefinition of coordinate time, one can  derive transformation equations  satisfying the above requirement. The related  clock synchronisation procedure is  referred to as the absolute synchronisation.\footnote{There are  more   advantages  of considering  that  formalism:  firstly -- it offers a wider framework  from which it is straightforward to move to STR; secondly -- its application to  the problem of localisation in quantum mechanics  allows to  define consistently covariant position and  spin  operators and formulate a covariant relativistic quantum mechanics free of inconsistencies~\cite{cite:spin-1}. }
Since the complete (covariant) formalism is somewhat complex, a derivation of the corresponding transformation equations is presented,  according to Rembieli\'nski and  W{\l}odarczyk~\cite{cite:RW}, in one spatial dimension and so are  the   subsequent discussions that constitute the esssence of the present article.

The aim of the  paper is threefold:  ($i$) present and discuss  basic properties of  transformation equations for space-time intervals  in  arbitrary synchronisation; ($ii$) present the absolute clock synchronisation procedure; ($iii$) demonstrate exceptional simplicity  in  explaining selected STR paradoxes on these grounds.
%re-examine selected STR paradoxes on theses grounds  and demonstrate  exceptional simplicity of these issues.

\section{Transformation equations in arbitrary synchronisation}\label{sec:1}
\paragraph{Synchronisation of distant clocks}
Assume two identical clocks (i.e. having the same duration of the unit time),  placed at  distant locations $A$ and $B$.
The  Einstein's procedure of  synchronising  these clocks is to send a light signal from $A$ to $B$, reflect it and receive at $A$. Clocks  are said to be synchronised when their readings are related by:
\begin{equation}\label{eq:tAtB}
t_{\scriptscriptstyle{E}}(B) = t_{\scriptscriptstyle{E}}(A) +  \frac{1}{2} \Delta t_{\scriptscriptstyle{ABA}},
\end{equation}
where $\Delta t_{\scriptscriptstyle{ABA}}$ denotes the time of flight over the closed path $ABA$, measured  at $A$; subscript $E$  identifies  the Einstein's synchronisation. A generalisation of   the clock synchronisation relation was proposed by Reichenbach~\cite{cite:Reichenbach-1}. It  consists in  replacing the factor $1/2$ in~(\ref{eq:tAtB}) with  a parameter  $\varepsilon_{\scriptscriptstyle{R}}$,  called the Reichenbach coefficient, leading to the following relation, below referred to as arbitrary synchronisation:
\begin{equation}\label{eq:tAtB-arb}
t_{\varepsilon_{\scriptscriptstyle{R}}}(B) = t_{\varepsilon_{\scriptscriptstyle{R}}}(A) + \varepsilon_{\scriptscriptstyle{R}} \Delta t_{\scriptscriptstyle{ABA}},
\end{equation}
where $0<\varepsilon_{\scriptscriptstyle{R}}<1$.  Merging~(\ref{eq:tAtB}) and~(\ref{eq:tAtB-arb}) leads to  the following relation between space-time intervals in the Einstein's and arbitrary synchronisations:
\begin{equation}
\Delta t_{\scriptscriptstyle{E}} = \Delta t_{\varepsilon_{\scriptscriptstyle{R}}} + (1-2\varepsilon_{\scriptscriptstyle{R}})\Delta x / c,
\end{equation}
where $\Delta t_{{\scriptscriptstyle{E}}} = t_{{\scriptscriptstyle{E}}}(B) - t_{{\scriptscriptstyle{E}}}(A)$,  $\Delta t_{\varepsilon_{\scriptscriptstyle{R}}} = t_{\varepsilon_{\scriptscriptstyle{R}}}(B) - t_{\varepsilon_{\scriptscriptstyle{R}}}(A)$, $\Delta x$ is the distance between $A$ and $B$ and $c$ is the average velocity of light over a closed path.
Introducing  a new synchronisation coefficient for convenience,
$\varepsilon = 1-2\varepsilon_{\scriptscriptstyle{R}}$, where $-1 <\varepsilon<1$,    one finally obtains:
\begin{equation}\label{eq:syn-11}
\Delta t_{\scriptscriptstyle{E}} = \Delta t_{\varepsilon}  +  \varepsilon\frac{\Delta x}{c}.
\end{equation}
Since space intervals  are measured  relative to a unit length (e.g. using  a ruler), they are  synchronisation independent, i.e. $\Delta x_{\scriptscriptstyle{E}} = \Delta x_{\varepsilon} \equiv \Delta x$.
%Moreover, the time lapse at any space point remains synchronisation independent  so the change of synchronisation scheme  is equivalent to  shifting %the origin of the time  axis ($t=0$) at each  point of space by the  value given by~(\ref{eq:syn-11}). ?????????

It follows from~(\ref{eq:syn-11}) that  relations between velocities in the Einstein's and arbitrary synchronisations,   $v_{\scriptscriptstyle{E}} = \Delta x / \Delta t_{\scriptscriptstyle{E}}$  and  $v_{\varepsilon} = \Delta x / \Delta t_{\varepsilon}$, respectively, read:
\begin{equation}\label{eq:vEveps}
v_{\scriptscriptstyle{E}} =  \frac{v_{\varepsilon}}{1 + \varepsilon v_{\varepsilon}/c}\;,  \qquad  v_{\varepsilon} =  \frac{v_{\scriptscriptstyle{E}}}{1 - \varepsilon v_{\scriptscriptstyle{E}}/c}.
\end{equation}
Note that changing sign of  $v_{\scriptscriptstyle{E}}$ does not imply  the same  for  $v_{\varepsilon}$ -- a property  reflecting  breaking of the reciprocity principle which  is   valid  in the Einstein's synchronisation only. Given two inertial reference frames, $\U$ and $\Uprime$,  if the velocity  of $\Uprime$ w.r.t. $\U$ equals  $+V_{\scriptscriptstyle{E}}$ in the Einstein's synchronisation and  $V_{\varepsilon}^{+}$ in arbitrary synchronisation,  then  the velocity  of $\U$ w.r.t. $\Uprime$ equals $-V_{\scriptscriptstyle{E}}$ in the Einstein's synchronisation and, following  from~(\ref{eq:vEveps}), $V_{\varepsilon}^{-} = - V_{\varepsilon}^{+}/ (1 + 2\varepsilon V_{\varepsilon}^{+}/c)$ in  arbitrary synchronisation.

\paragraph{One-way velocity of light}
It is straightforward to show, using~(\ref{eq:syn-11}), that the one-way velocity of light from $A$ to $B$ or from $B$ to $A$ in arbitrary synchronisation is given by, respectively:
\begin{equation}\label{eq:one-way-22}
c_{AB} = \frac{c}{1-\varepsilon}\;, \qquad\qquad c_{BA} = \frac{c}{1+\varepsilon}.
\end{equation}
As can be readily verified, the harmonic  average over a closed path  equals $c$, i.e. is  independent of synchronisation convention.
\paragraph{Transformation equations}
In order to  derive  transformation equations   for space-time intervals in  arbitrary synchronisation, $\Delta t_{\varepsilon}$ and $\Delta x$, between two inertial reference frames, $\U$ and $\Uprime$,  one inserts~(\ref{eq:syn-11}) into  the standard Lorentz transformation equations:
\begin{equation}\label{eq:Lorentz}
\left ( \begin{array}{cc}
c\Delta t^{\prime}_{\scriptscriptstyle{E}}\\
\Delta x^{\prime} \end{array} \right )
= \gamma_{\scriptscriptstyle{E}} \left(
\begin{array}{cc}
1 &  -V_{\scriptscriptstyle{E}}/c  \\
-V_{\scriptscriptstyle{E}}/c & \quad  1 \\
\end{array} \right)
\left ( \begin{array}{c}
c\Delta t_{\scriptscriptstyle{E}}\\
\Delta x \end{array} \right ),
\end{equation}
where $V_{\scriptscriptstyle{E}}$ denotes the velocity of $\Uprime$  w.r.t. $\U$ and $\gamma_{\scriptscriptstyle{E}}= 1/ \sqrt{1 - V_{\scriptscriptstyle{E}}^2/c^2}$ is the standard Lorentz factor. Taking into account that, in general,   the synchronisation coefficient $\varepsilon$  transforms too,   the relation~(\ref{eq:syn-11}) in the  reference frame $\Uprime$  reads: $\Delta t_{\scriptscriptstyle{E}}^{\prime} = \Delta t^{\prime}_{\varepsilon^{\prime}}  +  \varepsilon^{\prime} \Delta x^{\prime}/c$. One thus  obtains the following  relation of  time and space  intervals in arbitrary synchronisations:
\begin{subequations}\label{eq:2arbitrary}
\begin{equation}\label{eq:ARB}
\left ( \begin{array}{cc}
c\Delta t^{\prime}_{\varepsilon^{\prime}}\\
\\
\Delta x^{\prime} \end{array} \right )
= \gamma ({\varepsilon}) \left(
\begin{array}{cc}
1 + (\varepsilon + \varepsilon^{\prime})V_{\varepsilon}/c  & \qquad  \varepsilon - \varepsilon^{\prime} +  (\varepsilon^2 - 1) V_{\varepsilon}/c  \\
\\
-V_{\varepsilon}/c & 1 \\
\end{array} \right)
\left ( \begin{array}{c}
c\Delta t_{\varepsilon}\\
\\
\Delta x \end{array} \right ),
\end{equation}
while the  inverse transformation  becomes:
\begin{equation}\label{eq:ARB-inv}
\left ( \begin{array}{cc}
c\Delta t_{\varepsilon}\\
\\
\Delta x \end{array} \right )
= \gamma ({\varepsilon}) \left(
\begin{array}{cc}
1   & \quad\quad  - \Big ( \varepsilon - \varepsilon^{\prime} +  (\varepsilon^2 - 1) V_{\varepsilon}/c \Big ) \\
\\
V_{\varepsilon}/c  & \qquad 1 + (\varepsilon + \varepsilon^{\prime})V_{\varepsilon}/c\\
\end{array} \right)
\left ( \begin{array}{c}
c\Delta t^{\prime}_{\varepsilon^{\prime}}\\
\\
\Delta x^{\prime} \end{array} \right ),
\end{equation}
\end{subequations}
where
\begin{equation}\label{eq:gammaeps}
\gamma (\varepsilon) =  \frac{1} {\sqrt{(1 + \varepsilon V_{\varepsilon}/c)^2 - (V_{\varepsilon}/c)^2}}.
\end{equation}
Strictly speaking, the above do not yet constitute a complete set of  transformation equations  since  those  should comprise  a transformation law  for synchronisation coefficients, too.
The standard Lorentz transformation can be recovered  from~(\ref{eq:2arbitrary}) by putting $\varepsilon = \varepsilon^{\prime}=0$, i.e.  adopting the  Einstein's  synchronisation in both reference frames.
The above  equations  are, in general,  not reciprocal, which reflects  breaking of the relativity principle in arbitrary synchronisation. Reciprocity is restored  if $\varepsilon = - \varepsilon^{\prime}$ which comprises  the Einstein's synchronisation, $\varepsilon = \varepsilon^{\prime}=0$,   as well as  in the Galilean limit, $c\rightarrow \infty$.

\paragraph{Measurement of time intervals and length}
Let two observers be stationary in $\U$ and $\Uprime$ ($\Delta x=0$ and $\Delta x^{\prime}=0$, respectively). According to~(\ref{eq:2arbitrary}), the  former will state  that time flows at the following  rate  for the  latter  w.r.t.  his rate, $\Delta t^{\prime}_{\varepsilon^{\prime}}$:
\begin{subequations}
\begin{equation}\label{eq:timeflow-a}
\Delta t^{\prime}_{\varepsilon^{\prime}} = \gamma (\varepsilon)\Big ( 1 + (\varepsilon + \varepsilon^{\prime})V_{\varepsilon}/c \Big ) \Delta t_{\varepsilon}
\end{equation}
while the latter  will state  that time flows at the following  rate  for the former  w.r.t.  his rate, $\Delta t_{\varepsilon}^{\prime}$:
\begin{equation}\label{eq:timeflow-b}
\Delta t_{\varepsilon} = \gamma (\varepsilon)  \Delta t^{\prime}_{\varepsilon^{\prime}}.
\end{equation}
\end{subequations}

Let an object of length $L$ be positioned in $\U$  along the $x$ axis, the length being  the  distance between its  ends measured simultaneously in a given  reference frame.  Thus the  length   in $\Uprime$, $L^{\prime} \equiv \Delta x^{\prime}$,   equals the value  of the $\Delta x^{\prime}$ space interval calculated   under the condition of simultaneity in $\Uprime$ ($\Delta t^{\prime}_{\varepsilon^{\prime}} = 0$). Applying this condition to~(\ref{eq:ARB}) and eliminating  $\Delta t_{\varepsilon}$,   one obtains the following expression for the length of the object in $\Uprime$ in arbitrary synchronisation:
\begin{equation}\label{eq:length-arb_syn}
L^{\prime}_{\varepsilon^{\prime}} = \frac{L_{\varepsilon}}{\gamma (\varepsilon) \big (1+(\varepsilon + \varepsilon^{\prime})V_{\varepsilon}/c \big )}.
\end{equation}
Depending on the values or functional forms of the synchronisation coefficients (e.g. dependence on $V_{\varepsilon}$),  the length in $\Uprime$ may be unaltered, contracted or  elongated as compared to the length in $\U$.
Applying the  limit $c\rightarrow \infty$ in~(\ref{eq:length-arb_syn}) leads to  the result for the Galilean transformation,  $L^{\prime}_{\varepsilon^{\prime}} = L$. If  $\varepsilon = \varepsilon^{\prime} = 0$, the expression for the length contraction in STR, $L^{\prime} = L/\gamma_{\scriptscriptstyle{E}}$, is restored.

\section{Absolute synchronisation and a  preferred reference frame}\label{sec:2}

Among possible synchronisation schemes  one can distinguish such  that satisfies the requirement  of absolute  simultaneity, i.e. fulfilling the following proportionality:  $\Delta t^{\prime}_{\varepsilon} \sim  \Delta t_{\varepsilon}$.  This scheme   will be called the absolute synchronisation.
The corresponding transformation equations are  easily derived under the condition that the spatial  component  does not participate in the transformation of the time component. This is achieved by imposing  the following  requirement  in~(\ref{eq:2arbitrary}):
\begin{equation}\label{eq:cond-1}
\varepsilon - \varepsilon^{\prime} - (1 - \varepsilon \varepsilon^{\prime}) \;\frac{V_{\varepsilon}/c}{1 + \varepsilon V_{\varepsilon}/c}  = 0,
\end{equation}
which thereby enables to write the transformation law for the synchronisation coefficients.
%which thereby enables  to relate the synchronisation  coefficients  $\varepsilon$ and $\varepsilon^{\prime}$.
In consequence,  one obtains  the following transformation  equations between $\U$ and $\Uprime$ in  the absolute synchronisation (in what follows,  quantities  in this synchronisation are not marked by subscripts):
\begin{subequations}\label{subeq:pref-transf}
\begin{equation}\label{eq:preferred-transf-1}
\left ( \begin{array}{cc}
c \Delta t^{\prime}\\
\Delta x^{\prime} \end{array} \right )
= \left(
\begin{array}{cc}
1/ \gamma (\varepsilon) &  0  \\
%\hline
-\gamma (\varepsilon)\,V/c & \quad  \gamma (\varepsilon) \\
\end{array} \right)
\left ( \begin{array}{c}
c\Delta t\\
\Delta x \end{array} \right ),
\end{equation}
\begin{equation}\label{eq:preferred-transf-c}
\varepsilon^{\prime} = \varepsilon - (1-\varepsilon^2)V/c.
\end{equation}
\end{subequations}

The condition  $\varepsilon = 0$  marks a class of reference frames in which the  Einstein's synchronisation is valid and, according to~(\ref{eq:one-way-22}),  the one-way velocity of light equals $c$ in both directions of the $x$ axis.
One has  freedom  to select one reference frame belonging to this class and assign it the  status of  the  preferred  frame ($\UPF$).
It is  implicitly assumed there exist  no  physical  phenomena  in favour  of  a given  particular choice.
If one considers  motion of  bodies  with velocities smaller than $c$,  as  this is done  in the present paper, it is a matter of indifference which reference frame will be named  $\UPF$. However, once  a preferred frame  has been  chosen, the relativity principle in the standard  formulation  is broken in consequence. One can  formulate instead a  principle stating that any inertial reference frame  may be assumed the preferred frame (provided  the clocks in this reference frame are  synchronised according  to the Einstein's convention).

Since there is  freedom of choosing the preferred frame,  it is convenient   to   identify  one of the reference frames, involved in the formulation of STR problems,  with  $\UPF$ (e.g. $\U \equiv \UPF$),  subsequently  using   the tilde sign  to mark the  corresponding quantities.  Let the reference frame $\Uprime$ move with velocity $\widetilde{V}$ w.r.t. $\UPF$.
The following applies in  $\Uprime$: ($i$) inserting $\varepsilon=0$ in~(\ref{eq:preferred-transf-c}) leads to the  solution:  $\varepsilon^{\prime} = - \widetilde{V}/c$  which  defines the procedure of  absolute clock synchronisation in the moving frame;
%according to the relation: $t^{\prime}(B) = t^{\prime}(A) + \varepsilon^{\prime} \Delta t^{\prime}_{ABA}$;
($ii$) it follows from~(\ref{eq:one-way-22}) that the one-way velocity of light  is  direction dependent in $\Uprime$:
\begin{equation}\label{eq:c_plus_minus}
c_+^{\,\prime} = c/(1+\widetilde{V}/c)\;, \qquad c_-^{\,\prime} =  c/(1-\widetilde{V}/c),
\end{equation}
where $c^{\,\prime}_+$  and  $c^{\,\prime}_- $  are velocities  of light in the positive and negative direction of the $x^{\prime}$ axis, respectively. The  condition that the average velocity over a closed path equals $c$ is satisfied:  $ \frac{1}{2}(1/c^{\,\prime}_+ + 1/c^{\,\prime}_-) = 1/c $.

Transformation equations~(\ref{eq:preferred-transf-1})  between $\UPF$ and a moving frame $\Uprime$ reduce to a simple form:
\begin{subequations}\label{sub:PF}
\begin{equation}\label{eq:PF-7}
\left ( \begin{array}{cc}
c\Delta t^{\prime}\\
\Delta x^{\prime} \end{array} \right )
= \left(
\begin{array}{cc}
1/ \gamma_0 &  0  \\
%\hline
-\gamma_0\,\widetilde{V}/c & \quad  \gamma_0 \\
\end{array} \right)
\left ( \begin{array}{c}
c \Delta \widetilde{t}\\
\Delta \widetilde{x} \end{array} \right ),
\end{equation}
and
\begin{equation}\label{eq:PF-7-inv}
\left ( \begin{array}{cc}
c \Delta \widetilde{t}\\
\Delta \widetilde{x} \end{array} \right )
= \left(
\begin{array}{cc}
 \gamma_0 &  0  \\
%\hline
\gamma_0\,\widetilde{V}/c& \quad  1/ \gamma_0 \\
\end{array} \right)
\left ( \begin{array}{c}
c \Delta t^{\prime}\\
\Delta x^{\prime} \end{array} \right ),
\end{equation}
\end{subequations}
where $\gamma_0 = 1/\sqrt{1-(\widetilde{V}/c)^2}$.

Equation~(\ref{eq:cond-1}) has also one  specific  solution, $\varepsilon=1$ and $\varepsilon^{\prime}=1$,  which corresponds to   $\varepsilon_{\scriptscriptstyle{R}} = 0$ and $\varepsilon^{\prime}_{\scriptscriptstyle{R}} = 0$ in the synchronisation relation~(\ref{eq:tAtB-arb}). This case  corresponds to the instantaneous synchronisation or, equivalently,  absolute coordinate time, $t(B) = t(A)$. Equations~(\ref{eq:preferred-transf-1}), in the limit $c\rightarrow \infty$, describe the Galilean transformation in that case.

%However,  if superluminal particles (tachyons) existed,  a  distinguished --  preferred -- frame of reference must exist in the Universe. If so,  %absolute (instantaneous) synchronisation of distant clocks might be  achieved   with arbitrary precision using a tachyon of velocity correspondingly %large (approaching infinity). Transformation~(\ref{eq:preferred-transf-1}) is also applicable for space-like trajectories  and does not  lead to %causality violation.
%unlike in the case of the standard Lorentz transformation~(\ref{eq:Lorentz}).

\section{Paradoxes}\label{sec:3}

\subsection{The ladder--barn  paradox}\label{subsec:barn}

\paragraph{Einstein's synchronisation}
A ladder ($\Uprime$)  of  length $L$  is  moving through a barn ($\U$) of length $l<L$, entering  through the front door and leaving through the back door.  An  observer in the barn would state that, due to the Lorentz contraction,  the ladder can be fit in  the barn  instantaneously if its   velocity  is such  that $\gamma_{\scriptscriptstyle{E}} > L/l$, where $\gamma_{\scriptscriptstyle{E}}$ is the standard Lorentz factor. When the end of the ladder enters the front door of the barn, both  doors of the barn are closed simultaneously  to mark the  fact that the barn entirely contains the ladder.
%\footnote{The back door is   immediately opened again to let the ladder leave the barn.}
An observer on the ladder would state however that the ladder cannot be fit in the barn because the  length of the barn, owing  to the Lorentz contraction, is yet smaller: $l/\gamma_{\scriptscriptstyle{E}} <l<L$. The  apparent paradox consists in that simultaneous closing  the front and the back door of the barn are not simultaneous events in the reference frame of the ladder: the back  barn door closes  earlier than the front of the ladder leaves the barn while the front door closes just as the end of the ladder enters the barn.
Explanation of the problem in $\U$ focuses on the notion of the length  while understanding it  in  $\Uprime$  does not explicitly  involve lengths  but lies in  non-simultaneity.
%Thus for an observer on the ladder, it is   at no time  contained between the front and the back door of the barn. The designation "contained"  is %applicable to the barn only. Nevertheless, the length of the barn, if measured according to rules, would come out contracted.

\paragraph{Absolute  synchronisation}
%The  explanation  of  this thought experiment  in the absolute synchronisation is straightforward.
Making use of the freedom of choosing  a  preferred frame, assume that the barn  is identified with  $\UPF$. Justification of this choice follows from the above reasoning since the problem of fitting the ladder in the barn is well defined in the reference frame of the barn.
According to~(\ref{sub:PF}), an observer in  the barn   would measure  the length of the ladder ($\Delta \widetilde{t} = 0 $) as $\widetilde{L}=L / \gamma_0 $, i.e. the ladder will be contracted and will fit  in  the barn if $\gamma_0 \geq L/l$.  The observer on the ladder  would state that the length of the barn, $l^{\prime}$, is elongated  ($\Delta t^{\prime} = 0 $): $l^{\prime}=\gamma_0 l$.   Owing to absolute simultaneity, both observers perform  measurements of the length and both will  obtain the same condition for the ladder to fit in the barn.  {\it Ergo},  one avoids the  paradox when  this problem is  described   in the  absolute  synchronisation.
On the other hand, if one associates  the preferred frame with the ladder,  the description of the entire phenomenon would be  similar to that for   an  observer on the ladder using  the Einstein's synchronisation. In the absolute synchronisation, the  paradox does not appear also in this case since  both observers,  on the ladder and  in the barn, would agree that the barn becomes  contracted and the ladder  elongated  so  the ladder  cannot be fit in  the barn.  In  both reference frames, the  front and the  back  doors would be opened  non-simultaneously allowing the ladder to pass through  the barn.

\subsection{The  twin paradox}\label{subsec:twin}
\paragraph{Einstein's synchronisation}
Given twins in uniform relative motion, it follows from STR that each one would  state that  the  proper time  flows  at  a lower rate   for the other twin.
%These apparently contradictive -- paradoxical --  predictions,  result  from the Lorentz transformation,  and are  explicit manifestation   of the %relativity principle.
Verification of  respective  proper time intervals  elapsed from the beginning of the journey   may   be  possible only when the twins meet again and compare clock readings. If the formulation of the problem possesses a  kinematical asymmetry, such that e.g. one  twin turns around (travelling twin) to catch up with  the other who keeps moving  uniformly, then their clock readings  would be  different when eventually compared side-by-side. The travelling  twin  will be younger because the time interval elapsed in his reference frame during his journey will be smaller by the  standard $\gamma_{\scriptscriptstyle{E}}$ factor, assuming the simplest case  that the velocities of his motion,  w.r.t. the other twin,  in both directions had  equal values.

\paragraph{Absolute  synchronisation}
Assume  one twin  at rest in  $\UPF$  while   $\Uprime$ assigned to the other (travelling) twin. Time intervals elapsed in the reference frames of the  travelling twin and  the twin at rest  are related as follows,  according to~(\ref{eq:PF-7}): $\Delta t^{\prime}= \Delta  \widetilde{t}/\gamma_0$ while  according to~(\ref{eq:PF-7-inv}):  $\Delta \widetilde{t} = \gamma_0 \Delta t^{\prime}$ (this is valid for both legs of the journey).
Since these relations are equivalent, both twins agree already during the travel  that time flows at a  lower rate for the  travelling twin than for the  one at rest in $\UPF$. {\it Ergo}, there are  no paradoxical aspects  if  the problem is considered in the absolute synchronisation.

\subsection{The Bell's spaceship paradox}\label{subsec:spaceship}

\paragraph{Formulation}
The paradox was originally formulated by E.~Dewan  and M.~Beran in 1959~\cite{cite:Dewan-Beran}. It is presented  in short below,   following  J.~S.~Bell~\cite{cite:Bell-spaceship}. Two spaceships at rest in a given reference frame, $\U$, separated by a   distance $L$,  are   connected  with  a  taut string.   Both spaceships  accelerate simultaneously in such a way that the distance between them, as viewed in  $\U$, remains constant and equal $L$. Will  the string  break at a certain moment during the acceleration?
The problem received attention by several authors.  Dewan and Beran, as well as Bell, judged that the string will break.  There were however also contrary  opinions~\cite{cite:Nawrocki} and replies~\cite{cite:Dewan-2} (see also~\cite{cite:Matsuda-Kinoshita}).\footnote{Divergencies in opinions among  physicists have been noted by Bell~\cite{cite:Wiki}.}

\paragraph{Einstein's synchronisation}
Associate $\Uprime$ with one of the spaceships and assume that it has  reached velocity $V_{\scriptscriptstyle{E}}$ after  certain time of acceleration.
%Measuring  the distance between the spaceships  means taking the difference of their positions  simultaneously. These events are not simultaneous in %$\U$: let their time separation be $\Delta t_{\scriptscriptstyle{E}}$  so  the space separation is  $\Delta x = L + V_{\scriptscriptstyle{E}} \Delta %t_{\scriptscriptstyle{E}}$.
It can be shown that the  distance between the spaceships in the co-moving frame, $L^{\prime}$,  has increased  during acceleration: $L^{\prime} =
\gamma_{\scriptscriptstyle{E}} L$. Since the string is attached  to both spaceships, it has to break  once  its elasticity limit  is   exceeded.
On the other hand,  as viewed in $\U$, the distance between the spaceships  remains the same  but the elasticity limit of the string undergoes Lorentz contraction,  as argued by Bell,  leading to the same conclusion.

\paragraph{Absolute  synchronisation}
Assume that  $\UPF$ is the reference frame w.r.t. which the spaceships were  accelerated.
Substituting $\Delta \widetilde{x} = L$ and requesting  $\Delta \widetilde{t} = 0$ in~(\ref{eq:PF-7}) leads directly to the  result $L^{\prime} = \gamma_0 L$. Since the condition $\Delta \widetilde{t} = 0$ implies $\Delta t^{\prime} = 0$ then  $L^{\prime}$ is the  measured  distance between the two spaceships in the moving frame  at any moment of the accelerated motion  and the string has to  break when its  elasticity limit is exceeded.
This limit corresponds to a certain characteristic length in the rest frame of the string ($\Uprime$). According to~(\ref{sub:PF}), this  length is contracted by the factor of $\gamma_0$  for an observer in $\UPF$ which makes  a consistent  explanation for the string to break.
Interpretation  of the Bell's spaceship paradox in the absolute synchronisation  is thus simple and non-controversial.

\section{Summary and conclusions}\label{sec:summary}

Procedure of clock synchronisation is a convention in STR. One is allowed to adopt a convention  suitable  for solving a given problem.
In the Einstein's synchronisation  simultaneity is relative, being sometimes a source of certain  difficulties in analysing STR issues.   It has been  shown that,  by choosing the absolute synchronisation and  including  the preferred frame in solving  STR paradoxes,  it is straightforward to  explain  them   without analysing  simultaneity. The paradoxical features do not appear in consequence of these choices and, of course,  lead to identical results as in  the Einstein's synchronisation.

\vspace{1.0cm}

\noindent
{\bf Acknowledgements}

We are indebted to J.~Rembieli\'nski for numerous discussions of the preferred frame mechanics.

\end{document}